\documentclass[aps,prl,twocolumn,superscriptaddress,nobibnotes]{revtex4}

\usepackage{graphicx}
\newcommand{\ket}[1]{\,| \hspace{0.5ex} #1 \rangle}

\begin{document}

\title{Ultrafast all-optical switching by single photons}


\author{Thomas Volz}\thanks{These authors contributed equally to this work.}
\affiliation{Institute of Quantum Electronics, ETH Zurich, 8093
Zurich, Switzerland}
\author{Andreas Reinhard}\thanks{These authors contributed equally to this work.}
\affiliation{Institute of Quantum Electronics, ETH Zurich, 8093
Zurich, Switzerland}
\author{Martin Winger}
\affiliation{Institute of Quantum Electronics, ETH Zurich, 8093
Zurich, Switzerland}
\author{Antonio Badolato}
\affiliation{Department of Physics and Astronomy, University of
Rochester, Rochester, NY 14627, USA}
\author{Kevin J.\ Hennessy}
\affiliation{Institute of Quantum Electronics, ETH Zurich, 8093
Zurich, Switzerland}
\author{Evelyn L.\ Hu}
\affiliation{School of Engineering and Applied Physics, Harvard
University, Cambridge, Massachusetts 02138, USA}
\author{Ata\c c Imamo\u glu}
\affiliation{Institute of Quantum Electronics, ETH Zurich, 8093
Zurich, Switzerland}

\maketitle






{\bf An outstanding goal in quantum optics is the realization of
fast optical non-linearities at the single-photon level. Such
non-linearities would allow for the realization of optical devices
with new functionalities such as a single-photon switch/transistor
\cite{Chang:NatPhys07,Hwang:Nature09} or a controlled-phase gate
\cite{Turchette:PRL95}, which could form the basis of future quantum
optical technologies \cite{Obrien:NatPhot09}. While non-linear
optics effects at the single-emitter level have been demonstrated in
different systems, including atoms coupled to Fabry-Perot or
toroidal micro-cavities
\cite{Turchette:PRL95,Schuster:NatPhys08,Dayan:Science08},
super-conducting qubits in strip-line resonators
\cite{Fink:Nature08,Deppe:NatPhys08,Bishop:NatPhys09} or quantum
dots (QDs) in nano-cavities
\cite{Hennessy:Nature07,Srinivasan:Nature07,
Fushman:Science08,Kasprzak:NatMat10}, none of these experiments so
far has demonstrated single-photon switching on ultrafast
timescales. Here, we demonstrate that in a strongly coupled
QD-cavity system the presence of a single photon on one of the
fundamental polariton transitions can turn on light scattering on a
transition from the first to the second Jaynes-Cummings manifold
with a switching time of 20~ps \cite{Faraon:NatPhys08}. As an
additional device application, we use this non-linearity to
implement a single-photon pulse-correlator. Our QD-cavity system
could form the building-block of future high-bandwidth photonic
networks operating in the quantum regime
\cite{Faraon:APL07,Brossard:APL10,Bose:OptExp11}.}

\begin{figure} [h!]
\includegraphics[width= 80 mm]{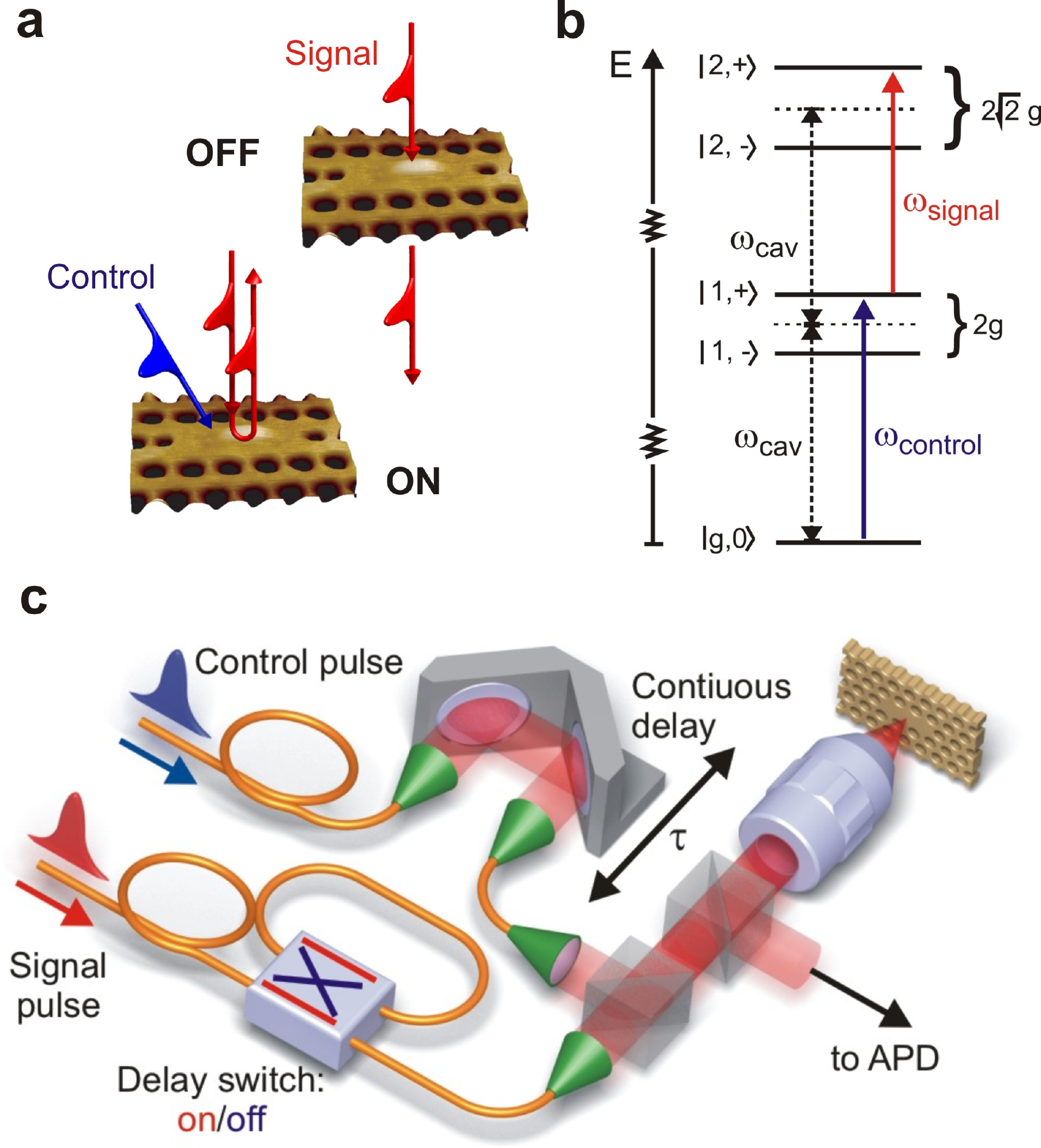}
\caption{{ \bf A single-photon all-optical switch.} {\bf a,} A
single control photon incident on the QD-cavity device determines
whether a signal photon of different colour is scattered. {\bf b,}
Energy-level diagram of the strongly coupled QD-cavity system up to
the second manifold of the anharmonic Jaynes-Cummings ladder. A
single control photon on the upper polariton transition (UP) to the
first manifold $\left(\ket{\mathrm{g},0} \rightarrow
\ket{1,+}\right)$ changes the scattering rate of a second signal
photon resonant with a transition from the first to the second
manifold $\left(\ket{1,+} \rightarrow \ket{2,+}\right)$. {\bf c,}
Setup for the demonstration of ultrafast single-photon switch
operation. The relative delay between signal and control pulses is
adjusted by a continuous delay stage. In addition, a discrete delay
line, corresponding to a time delay of $\approx 5$~ns, can be added
to the path of the signal pulse using a fibre switch. The photons
back-scattered from the QD-cavity system are detected by an
avalanche photodiode (APD) in single-photon counting mode.
\label{Fig1}}
\end{figure}

A single-photon switch is a device where a single optical gate (or
control) photon controls the propagation (or scattering) of incident
signal photons through non-linear optical interactions
\cite{Chang:NatPhys07}, as illustrated in Figure~1a. The requisite
large photon-photon interactions are provided by the basic system of
cavity quantum electrodynamics (QED) \cite{Mabuchi:Science02},
namely that of a single quantum emitter strongly coupled to a single
cavity mode: Figure~1b displays the first two rungs of the
corresponding energy diagram, the so-called Jaynes-Cummings ladder.
The anharmonicity of the Jaynes-Cummings ladder with the resulting
single-photon non-linearity is due to the quantized nature of the
radiation field and has been demonstrated for almost all cavity-QED
implementations either spectroscopically
\cite{Srinivasan:Nature07,Schuster:NatPhys08,Fink:Nature08,Deppe:NatPhys08,Fushman:Science08,Bishop:NatPhys09,Kasprzak:NatMat10}
or through photon-correlation measurements
\cite{Birnbaum:Nature05,Faraon:NatPhys08,Kubanek:PRL08,Dayan:Science08,
Lang:PRL11,Reinhard:arXiv11}. However, the speed of devices that are
based on single-photon non-linearities have so far not been
addressed. Since the ultimate switching times are limited by the
reciprocal emitter-cavity coupling strength, quantum dots strongly
coupled to nano-cavities
\cite{Yoshie:Nature04,Reithmaier:Nature04,Peter:PRL05,Hennessy:Nature07}
emerge as ideal candidates for the realization of ultrafast
single-photon non-linear devices, due to their record-high coupling
strengths.

\begin{figure*}
\includegraphics[width= 175 mm]{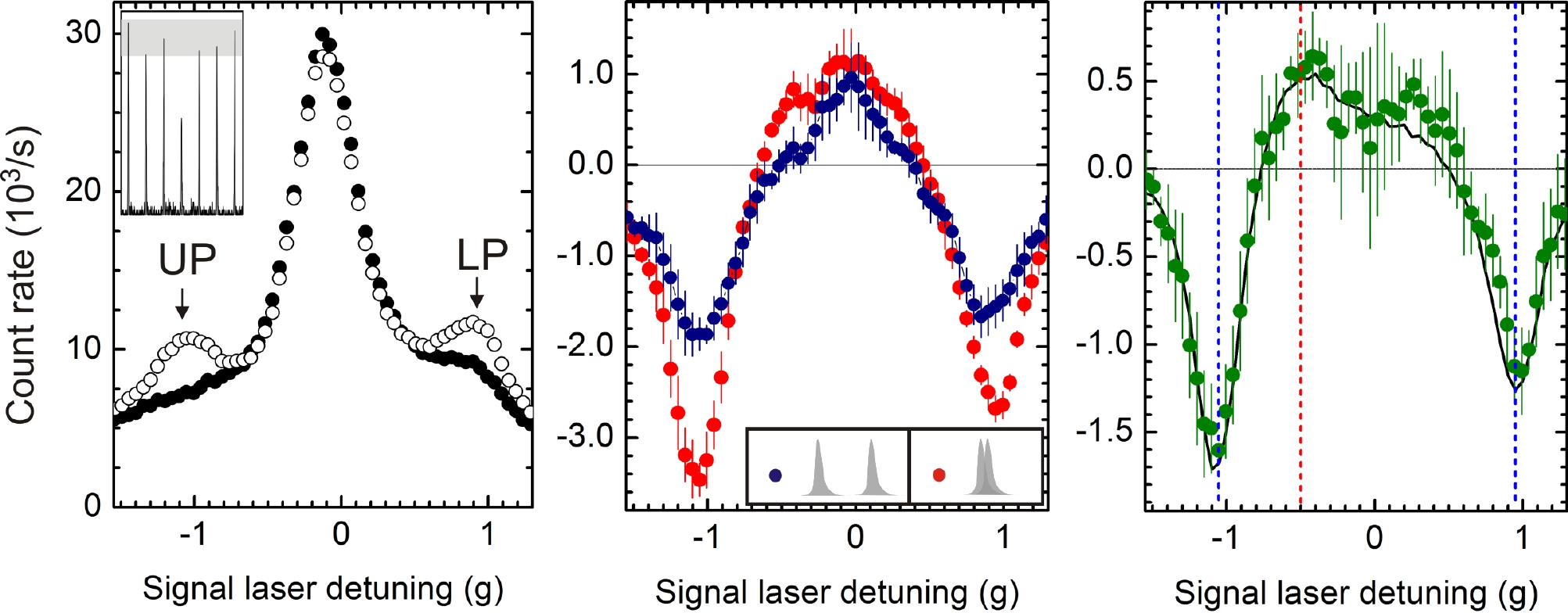}
\caption{{\bf Two-colour spectroscopy of the strongly coupled
QD-cavity device.} {\bf a,} System response with ($\bullet$) and
without ($\circ$) control laser present when scanning the signal
laser across the QD-cavity spectrum (for a cavity blue-detuning of
$\approx -0.1$g). Pulse durations of both control and signal laser
pulses were $86$~ps. Inset: autocorrelation histogram taken on the
upper polariton transition (pulse duration 72~ps) clearly
demonstrating photon blockade. {\bf b,} Non-linear behavior is
observed for a time delay between the control and the signal pulses
of 5~ns (blue data points) and $25$~ps (red data points). {\bf c,}
Subtracting the red and the blue data points of part (b) from each
other, we obtain the system non-linearity due to the Jaynes-Cummings
dynamics taking place on ultrashort timescales. In (b) and (c),
three adjacent data points were averaged. The black curve in (c) was
obtained from a Monte-Carlo wave-function simulation of the system
dynamics with the experimental parameters as input. \label{Fig2}}
\end{figure*}

Our device consists of an InAs/GaAs quantum dot positioned at an
electric-field maximum of a photonic crystal defect cavity in L3
geometry \cite{Hennessy:Nature07}. The QD-cavity system is deep in
the strong coupling regime: With a coherent coupling constant $g$ of
$141~\mathrm{\mu eV}$ and a quality factor of $Q\approx25~000$, i.e.
a cavity photon decay rate $\kappa$ corresponding to $53~\mathrm{\mu
eV}$, the figure of merit for the anharmonicity of the coupled
system, $g/\kappa \approx 2.7$, well exceeds the previously reported
values in literature. Resonant spectroscopy of the strongly coupled
system is performed using a crossed-polarization technique which
ensures efficient suppression of the excitation-laser light
back-reflected from the sample surface \cite{Reinhard:arXiv11}.
Ultrafast laser pulses with pulse durations between 33 and 86~ps are
derived from a mode-locked Ti:Sa laser (see Methods). The pulse
delay between control and signal pulse is adjusted by a motorized
delay stage (Fig.~1c).

We first carry out resonant spectroscopy of the strongly coupled
QD-cavity system by scanning the center frequency of the pulsed
laser across the polariton spectrum (see Methods) when QD and cavity
are very close to resonance. The result of a single scan is
displayed as open circles in Fig.~2a for a pulse duration of 86~ps
and an average signal power of 1~nW. As reported previously
\cite{Reinhard:arXiv11}, we employ an off-resonant re-pump laser at
a repetition rate of 1~MHz to partially counteract the laser-induced
quantum-dot blinking present in the system. Due to this blinking, we
obtain a three peak spectrum consisting of the upper and lower
polariton peaks (UP and LP) and the uncoupled cavity peak in the
middle (see Supplementary Material). In order to ensure that we work
in the single-photon regime, we carried out photon auto-correlation
measurements \cite{Reinhard:arXiv11} demonstrating photon
antibunching, i.e.\ photon blockade \cite{Birnbaum:Nature05}, on
both upper and lower polariton transitions to the first
Jaynes-Cummings manifold (see inset of Fig.~2a).

Next, we perform two-colour spectroscopy of the strongly coupled
QD-cavity system. We tune the control laser, with average power
2~nW, to the upper-polariton resonance and scan the signal laser
pulse across the spectrum. For this scan, the delay of the signal
pulse with respect to the control pulse was chosen to be 25~ps
corresponding to the polariton lifetime. The resulting spectrum is
again displayed in Fig.~2a: the filled (open) circles show the
system response with (without) control laser present. The difference
between the two data sets directly reflects the non-linear response
of the QD-cavity system, with the reduction of the polariton signal
being the most obvious effect of the presence of the control laser.
We plot this difference in Fig.~2b (red data points). In addition to
the fast photon-photon interactions of interest, the coupled system
also exhibits a slow non-linearity caused by the laser-induced QD
blinking discussed in the previous paragraph (see Supplementary
Material). To distinguish between the two effects, we acquire the
same spectra as in Fig.~2a but now with a time-delay of 5~ns between
the control and signal pulse which is much longer than the polariton
lifetime and the laser-pulse widths. The difference signal is again
displayed in Figure~2b as blue bullets. Subtracting the two
non-linear responses in Fig.~2b from each other yields the fast
non-linear optical response from the strongly coupled QD-cavity
system and is depicted in Fig.~2c: the data shows that the largest
non-linear effect occurs at the spectral position of the polaritons
(vertical blue lines). Here, the change in the scattering rate
induced by the control laser is negative ($\approx -15\%$) due to
saturation of the corresponding transitions. At the transition
wavelength from the first to the second manifold (vertical red
line), this change is positive ($\approx +6\%$) since the absorption
of the control photon enables the subsequent scattering of a signal
photon. For confirmation, we performed numerical simulations based
on a Monte-Carlo wave-function (MCWF) approach with the experimental
parameters as input and only the absolute amplitude of the
non-linear signal determined from a least-square fit to the data
(see Supplementary Material). The excellent agreement between theory
(black line) and experiment clearly demonstrates that the observed
positive non-linearity is indeed due to the two-colour transition to
the second Jaynes-Cummings manifold. We emphasize here, that due to
the finite linewidth of the coupled-system eigenstates as well as
the finite bandwidth of the laser pulses there is some overlap
between transitions from the first to the second and from the second
to the third manifold and so forth. Hence, we expect a
non-negligible contribution to the non-linear signal stemming from
states higher up in the JC ladder.

\begin{figure}
\includegraphics[width= 77 mm]{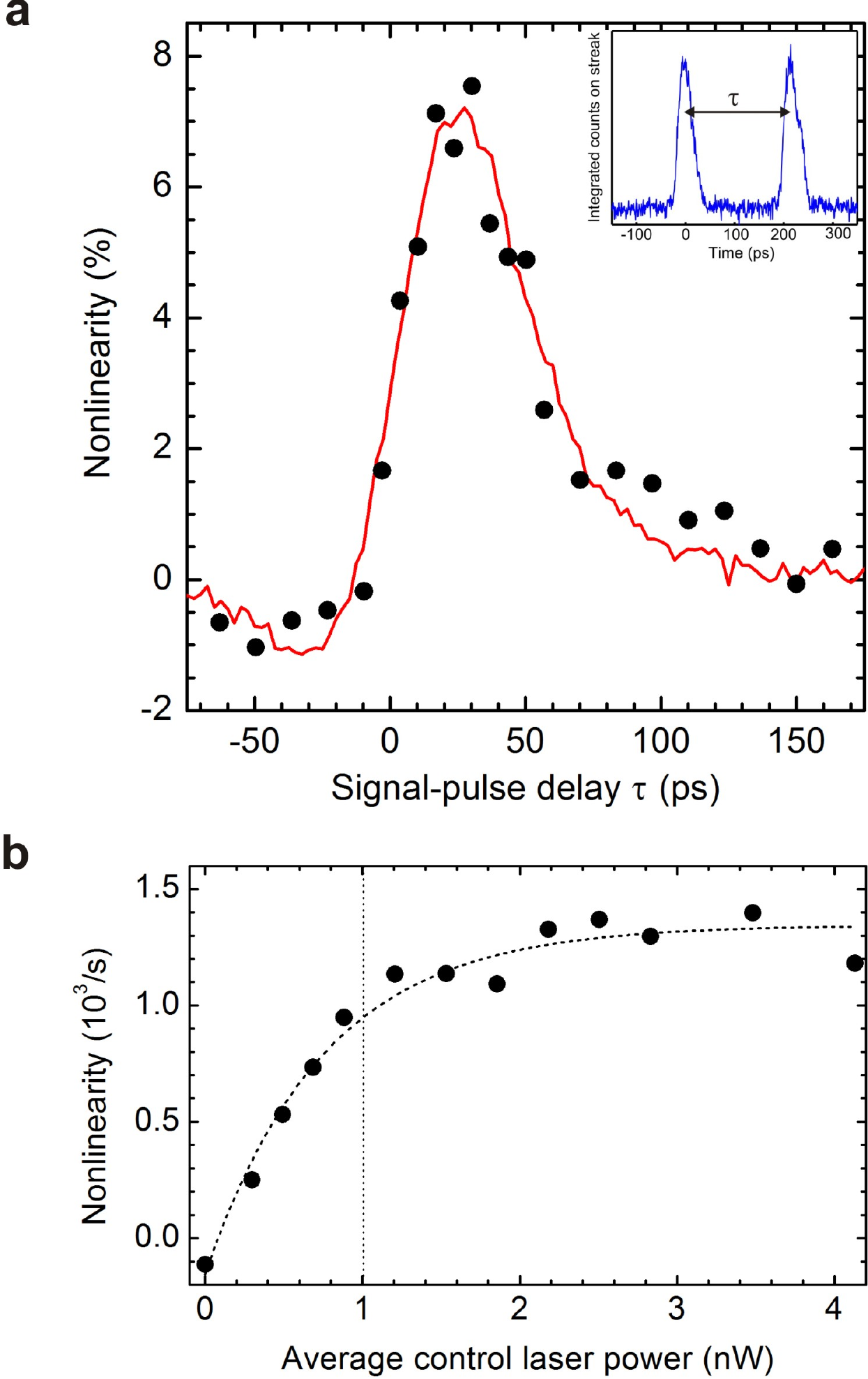}
\caption{{\bf Ultrafast non-linear response.} {\bf a,} Measured
non-linear response of the QD-cavity system (black points) as a
function of delay between control and signal laser pulses for a FWHM
of the pulses of 33~ps. Here, the non-linearity is given as relative
increase/decrease of the detected photon scattering rate compared to
the case without control laser. The sharp turn-on at zero time-delay
takes place over 20~ps -- demonstrating unprecedented switching
times in a non-linear photonic device. At positive time delays, when
the control photon arrives before the signal photon, the polariton
lifetime sets the relevant timescale. The red curve was obtained
from a numerical simulation using a Monte-Carlo wave-function
approach with the experimental parameters as input. {\bf b,}
Transfer characteristic of the single-photon switch. Here, the
non-linear signal was recorded as a function of control power for a
pulse delay of $25$~ps. For average powers larger than 1~nW, the
system saturates. The dashed line is a guide to the eye.
\label{Fig3}}
\end{figure}

In order to demonstrate ultrafast switching by single photons, i.e.\
conditional scattering of signal photons on ultrafast timescales, we
vary the delay between control and signal pulses while recording the
(positive) non-linearity for fixed laser-detunings. As depicted in
Fig.~1b, we choose the control pulse to be resonant with the
fundamental upper polariton and the signal pulse to be resonant with
the transition from the first to the second manifold. The result is
plotted in Figure~3a for a pulse duration of 33~ps; the black dots
depict experimental data, while the red curve was obtained from a
MCWF simulation with the absolute amplitude extracted from a
least-square fit to the data. Our single-photon switch exhibits both
ultra-fast turn-on and turn-off characteristics: The sharp turn-on
of the non-linear response of the QD-cavity system around zero time
delay reflects the response time of the system and clearly
demonstrates ultrafast switching operation of the strongly coupled
device: we find that the corresponding turn-on time (the time the
signal takes to rise from $10\%$ to $90\%$ of the maximum) is about
20~ps. At delay times longer than $\approx 30$~ps, the non-linear
signal of Fig.~3a exhibits a fast exponential decay corresponding to
the polariton lifetime which in turn determines the turn-off time of
our device. The asymmetry in the pulse-delay dependence is a direct
consequence of the cascaded-nature of the underlying two-photon
transition.

The ultimate switching time is in principle determined by the
anharmonicity of the Jaynes-Cummings spectrum, and is given by $((2-
\sqrt{2}) g)^{-1} \approx 20$~ps -- close to the switching time
observed in the experiment. Besides the switching speed, a quantity
of interest is the transfer characteristic, i.e.\ the output signal
as a function of control power which is plotted in Fig.~3b. Here, we
recorded the signal photons scattered from the system as a function
of input power of the control pulse. As expected, we first see a
linear increase with control power and then saturation when the mean
intra-cavity average photon number reaches $\approx 0.25$.

\begin{figure}
\includegraphics[width= 80 mm]{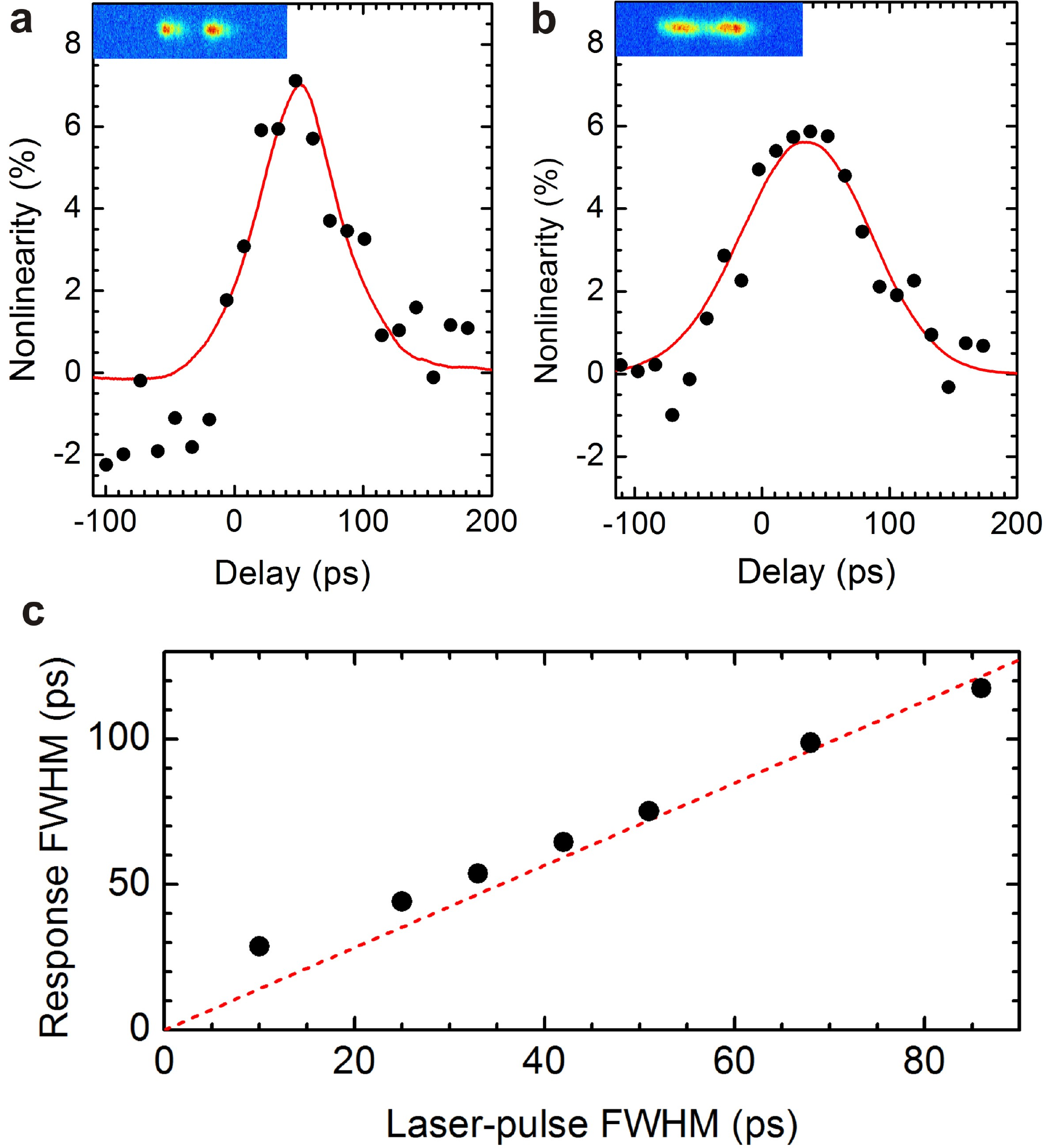}
\caption{{\bf Single-photon pulse-correlator.} The strength of the
non-linearity, as defined in the caption of Fig.~3, as a function of
delay for pulse durations of {\bf a,} 51~ps and {\bf b,} 86~ps. The
longer the pulse, the more symmetric the non-linear response. The
red curves are the numerical convolutions of the streak-camera
images shown as insets. The absolute amplitudes and the peak
positions of the convoluted signals were fit to the data. {\bf c,}
The simulated width ($\bullet$) of the non-linear system response
versus laser-pulse width. For pulse durations larger than 50~ps, the
simulated width of the non-linear signal approaches that of the
correlator width of the laser pulses (red dashed line). For very
short pulses, the polariton lifetime sets the lower limit for the
system response. \label{Fig4}}
\end{figure}

Besides the realization of single-photon switching, the strong
non-linearity of the QD-cavity system can be applied for measuring
pulse widths of ultrafast optical pulses down to the single-photon
level. This is demonstrated in Figure~4, where we map out the
non-linear system response as a function of pulse delay for pulse
durations of 51~ps and 86~ps. Since the pulse widths are
significantly longer than the polariton lifetime of 25~ps, the
system response is more symmetric than in the case of the 33~ps
pulses of Fig.~3a. We find good agreement between the
pulse-delay-dependent non-linearity and the numerical convolutions
of the independently obtained streak-camera images (red lines).
Figure~4c compares the FWHM obtained from Monte-Carlo simulations of
the non-linear system response with the FWHM of the incident
(Gaussian) laser pulses. The red dashed line corresponds to the
correlator width of the pulses. Above $50$~ps, the deviation of the
simulated width from this line is less than $5~\%$ -- in this range
our device works nicely as a single-photon pulse-correlator.
Finally, we remark that single-photon switch and/or pulse-correlator
operation can also be realized centering the signal pulse on the
other (lower) polariton transition, which yields a larger magnitude
for the non-linearity (see Fig.~2c and Supplementary Material).

A natural extension of our work would be the realization of a
single-photon transistor where the presence of a single
control-photon ($N_c = 1$) enables the scattering of $N_s \ge 2$
signal photons \cite{Chang:NatPhys07,Hwang:Nature09}. A simple
calculation shows (see Supplementary Material) that if pure
dephasing were absent, our QD-cavity device would exhibit a modest
gain of $G = N_s/N_c > 2$. While increasing the ratio $g/\kappa$
would already increase $G$, high-gain ($G\gg1$) transistors may be
realized in combination with EIT schemes
\cite{Imamoglu:PRL97,Muecke:Nature10,Hoi:PRL11,Suzuki:Science11}.
Another landmark achievement would be the demonstration of
preservation of quantum coherence during the non-linear interaction,
which could then pave the way for the realization of an ultrafast
controlled-phase gate between two single-photon pulses
\cite{Duan:PRL04}.

This work is supported by NCCR Quantum Photonics (NCCR QP), research
instrument of the Swiss National Science Foundation (SNSF) and an
ERC Advanced Investigator Grant (A.I.). The authors thank J.M.
Sanchez and U. Grob for assistance in the lab. The authors declare
that they have no competing financial interests. Correspondence and
requests for materials should be addressed to T.V. and A.I. (E-mail:
volz@phys.ethz.ch, imamoglu@phys.ethz.ch)

\section*{Methods}


\subsection*{Pulse preparation}

Both control and signal laser pulses are derived from the same
mode-locked Ti:Sa laser with a pulse repetition rate of $76.3$~MHz
and an intrinsic pulse width of a few picoseconds. The laser pulses
are sent through a grating spectrometer for frequency filtering and
split by a 50/50 beam splitter. Both beams are then coupled into
single-mode optical fibres. The resulting spectral width of the
pulses can be adjusted from 0.04~nm to 0.015~nm by an additional
slit in front of the spectrometer which determines the effective NA
of the spectrometer. The pulses are nearly Fourier-limited, thus we
can adjust the pulse duration from about 33~ps to 86~ps. We
mechanically tune the center frequency of the signal pulse by
employing a piezo-driven mirror holder in front of the fibre coupler
which enables coupling of different parts of the spectrum into the
fibre. The center frequency of the pulse is monitored using a
wavemeter, and a computer-controlled feedback loop allows for
tuning. The average power of both control and signal laser beams is
stabilized using acousto-optical modulators. The relative delay of
the two pulses is adjusted using a motorized delay stage. Pulse
shapes and delays are monitored by sending the light reflected from
the sample surface to a streak camera with approximately 4~ps time
resolution.

\subsection*{Extraction of the optical non-linearity}

When applying both a control pulse at time $t$ and a signal pulse at
time $t+\tau$ to the system, the time-integrated response can be
written as $N_{\mathrm{both~on}}(\tau) = N_{\mathrm{control}} +
N_{\mathrm{signal}} + N_{\mathrm{nl}}(\tau)$, where
$N_{\mathrm{control}}$ and $N_{\mathrm{signal}}$ denote the number
of scattered photons when only a control respectively signal laser
is applied. $N_{\mathrm{nl}}(\tau)$ is the total optical
non-linearity, quantified as the number of additional scattered
photons. Its origin is twofold: A fast ($\approx \mathrm{ps}$)
contribution arising from the anharmonicity of the Jaynes-Cummings
ladder (JC) and a slow ($\approx \mathrm{\mu s}$) contribution
stemming from charge blinking, thus $N_{\mathrm{nl}}(\tau) =
N_{\mathrm{nl,JC}}(\tau) + N_{\mathrm{nl,blinking}}(\tau)$. If we
choose an intermediate timescale $\tau_{\mathrm{int}}$ on the order
of ns, such that $\mathrm{ps} \ll \tau_{\mathrm{int}} \ll
\mathrm{\mu s}$, then $N_{\mathrm{nl,JC}}(\tau_{\mathrm{int}})
\approx 0$. For $\tau \ll \mathrm{\mu s}$,
$N_{\mathrm{nl,blinking}}(\tau_{\mathrm{int}}) \approx
N_{\mathrm{nl,blinking}}(\tau)$ and

\begin{equation}
\begin{array}{c}
N_{\mathrm{nl,JC}}(\tau) \approx N_{\mathrm{nl}}(\tau) - N_{\mathrm{nl}}(\tau_{\mathrm{int}}),\\
\\
\mathrm{where}~N_{\mathrm{nl}}(\tau) = N_{\mathrm{both~on}}(\tau) -
N_{\mathrm{control}} - N_{\mathrm{signal}}.
\end{array}
\end{equation}

In order to determine $N_{\mathrm{nl}}(\tau_{\mathrm{int}})$ we
choose $\tau_{\mathrm{int}} \approx 5~\mathrm{ns}$, by switching an
additional delay line into the path of the signal laser. In order to
eliminate long-time drifts we measure $N_{\mathrm{both~on}}(\tau)$,
$N_{\mathrm{control}}$ and $N_{\mathrm{signal}}$ simultaneously by
switching the control and signal lasers on/off with $5~\mathrm{kHz}$
and $10~\mathrm{kHz}$ respectively and sorting the output photons
accordingly.



\setcounter{figure}{0} \setcounter{equation}{0}
\renewcommand{\thefigure}{S\arabic{figure}}
\renewcommand{\theequation}{S\arabic{equation}}

\renewcommand{\thefigure}{S\arabic{figure}}
\renewcommand{\theequation}{S\arabic{equation}}

\clearpage

\section*{Supplementary Information}

\vspace{0.1cm}

\subsection*{Estimating typical blinking times}

As described in a previous publication \cite{Reinhard:arXiv11}, our
QD-cavity device exhibits significant blinking due to charging when
resonantly excited with laser light. For very low excitation powers,
the typical blinking time $\tau_\mathrm{b}$ (corresponding to the
lifetime of the neutral QD state in presence of laser light) was
determined from auto-correlation measurements using a single APD. We
find that $\tau_\mathrm{b}$ decreases linearly with increasing
average laser power. As a result, the polariton signals saturate and
the number of photons scattered on the bare cavity mode increases
quadratically with increasing power (see Fig.~S1a). In contrast to
the Jaynes-Cummings non-linearities, this non-linearity occurs on
the blinking timescale $\tau_\mathrm{b}$.

\begin{figure} [h!]
\includegraphics[width= 78 mm]{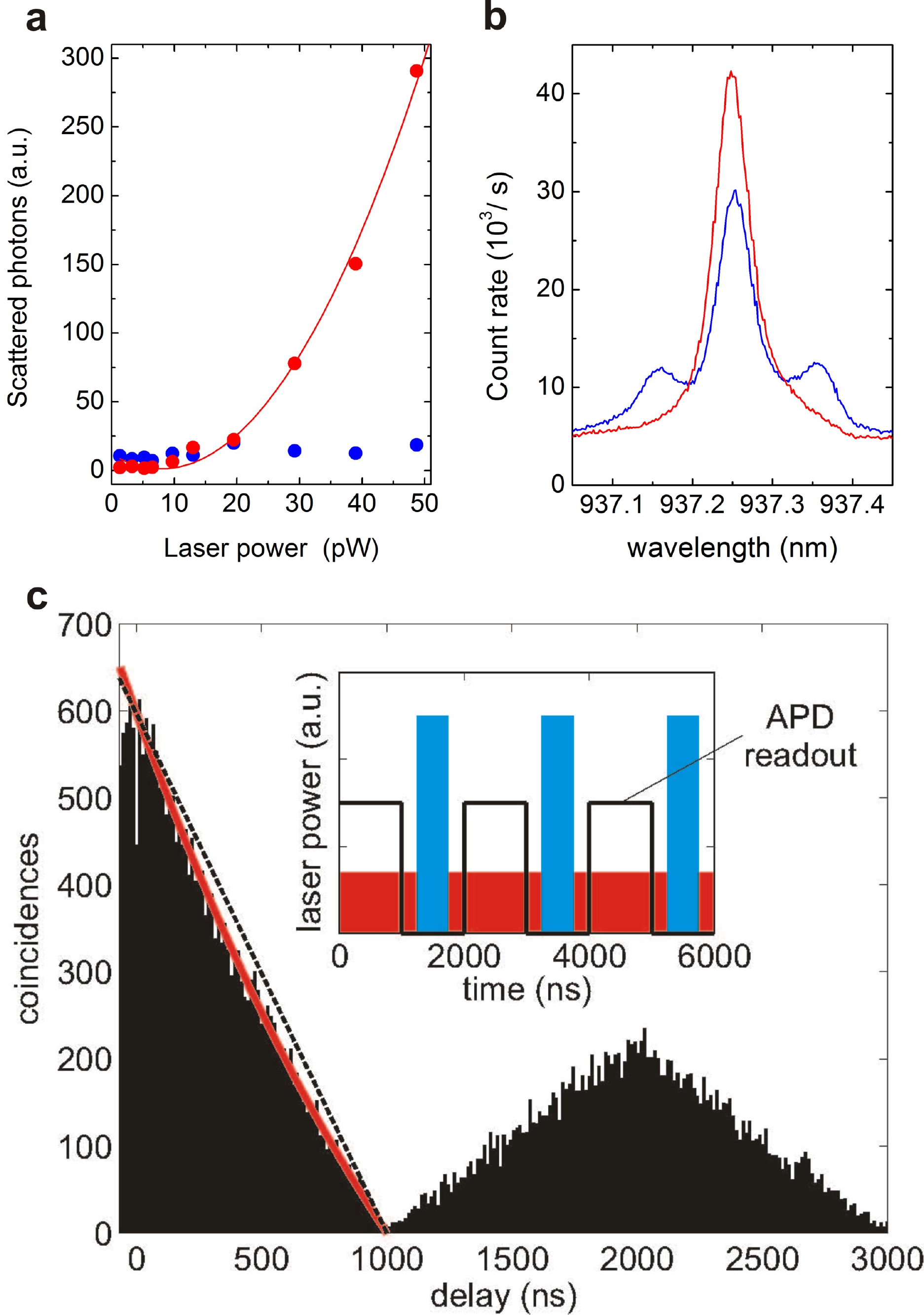}
\caption{{\bf Quantum dot blinking.} {\bf a,} Even at extremely low
laser powers, the signal from the QD-cavity device shows signs of
laser-induced blinking: the number of photons scattered from the
uncoupled cavity mode (red bullets) grows quadratically, while the
number of scattered photons on the polaritons (blue bullets)
saturates at very low levels. {\bf b,} A re-pump laser at 857~nm
\cite{Reinhard:arXiv11} counteracts the blinking and partially
recovers the polariton signal (blue line) at nW light levels.
Without re-pump laser, the polariton signals are vanishingly small
(red line). {\bf c,} Correlation histogram for pulsed excitation of
the upper polariton recorded in pulsed APD-readout mode (black
pulses in inset) with the re-pump laser present (blue pulses in
inset). The resonant laser was on permanently (red bar in the
inset). The red line on top of the histogram is a fit of the
slightly curved trailing slope of the center triangle, while the
dashed straight line is shown for reference. \label{FigS1}}
\end{figure}

We counteract the blinking effect by shining in an off-resonant
laser for re-pumping and thus partially restoring the polariton
signal (see Fig.~S1b). Since we find a typical blinking time on the
order of a few $\mu s$, the re-pump laser repetition frequency is on
the order of MHz. For determining $\tau_\mathrm{b}$ at nW light
levels, we record the photon auto-correlation function with a pulsed
laser resonant with a polariton transition, as described in
\cite{Reinhard:arXiv11}. Figure~S1c displays a typical
auto-correlation histogram taken on the upper polariton for an
excitation power of $2~\mathrm{nW}$ and a re-pump pulsing rate of
$1/T_{\mathrm{repump}}=0.5~\mathrm{MHz}$. As a consequence, the
auto-correlation function exhibits a periodic triangular shape on
the $\mathrm{\mu s}$ scale. The triangle at zero time delay is about
a factor of 3 larger than triangles at other times. This (classical)
bunching effect indicates that the system, on average, is in the
desired neutral ground state after every third re-pump pulse. The
detected polariton emission after a re-pump pulse is given by
$I\left(t\right)=I_{\mathrm{pol}}~e^{-t/\tau_{\mathrm{b}}}$, if
$0<t<T_{\mathrm{repump}}/2$ and $I\left(t\right) = 0$, if
$T_{\mathrm{repump}}/2<t<T_{\mathrm{repump}}$. The classical
auto-correlation function therefore reads as follows

\begin{equation}
\label{eq:g2}
\begin{array}{c}
G^{\left(2\right)}\left(\tilde{t}\right)=\int_{0}^{T_{\mathrm{repump}}}I\left(t\right)I\left(t+\tilde{t}\right)dt\\
=\frac{1}{2}I_{\mathrm{pol}}^{2}~\tau_{\mathrm{b}}~e^{-\tilde{t}/\tau_{\mathrm{b}}}\left(e^{-2\tilde{t}/\tau_{\mathrm{b}}}-e^{-T_{\mathrm{repump}}/\tau_{\mathrm{b}}}\right)\\
\mathrm{for}~ 0<\tilde{t}<T_{\mathrm{repump}}/2.
\end{array}
\end{equation}

We fit the trailing slope of the central triangle to the formula
above, resulting in a blinking time $\tau_{\mathrm{b}} = \left(6.2
\pm 1.2\right)\mathrm{\mu s} \gg \tau_{\mathrm{int}}$.

In the experiment described in the main text, we record the system
non-linearity for two different cases. In the first case, control
and signal pulses overlap in time ($\left|\tau\right| <
200~\mathrm{ps}$), in the second case, they are separated by several
ns ($\tau_{\mathrm{int}}\approx 5~\mathrm{ns}$). For the extraction
of the Jaynes-Cummings optical non-linearity, we calculate
$N_{\mathrm{nl,JC}}(\tau) = N_{\mathrm{nl}}(\tau) -
N_{\mathrm{nl}}(\tau_{\mathrm{int}})$ (see Methods, Eqn.~(1)), which
is based on the assumption that the non-linearity due to charge
blinking remains unchanged within time $\tau_{\mathrm{int}}$. As the
above analysis shows, this is justified, since
$\tau_{\mathrm{blinking}} \gg \tau_{\mathrm{int}}$.

\subsection*{Calculation of the optical non-linearity}

As described in the Methods section in the main text, we specify the
optical non-linearity as $N_{\mathrm{nl}}(\tau) =
N_{\mathrm{both~on}}(\tau) - N_{\mathrm{control}} -
N_{\mathrm{signal}}$. Here, $\tau$ is the delay of the signal pulse
with respect to the control pulse, and $N_{\mathrm{both~on}}(\tau)$,
$N_{\mathrm{control}}$ and $N_{\mathrm{signal}}$ are the number of
emitted photons per pulse (pulse pair) when either both laser pulses
together, only the control or only the signal laser are applied,
respectively. The photon number per pulse is given by
$N_{\mathrm{both~on}}(\tau) =
\int_{-T_{\mathrm{rep}}/2}^{T_{\mathrm{rep}}/2}{I\left(t\right)dt}$,
where $I\left(t\right)$ is the rate of emitted photons given by
$I\left(t\right)=\left<\mathrm{g}\right|\hat{C}^{\dagger}\left(t\right)\hat{C}\left(t\right)\left|\mathrm{g}\right>$,
where $\left|\mathrm{g}\right>$ denotes the system ground state,
before a laser pulse has excited the system. The system collapse
operator $\hat{C}\left(t\right)=\sqrt{\kappa}~\hat{a}\left(t\right)$
evolves according to the non-Hermitian effective Hamiltonian
$H_{\mathrm{eff}}\left(t\right)=H_{\mathrm{JC}}+H_{\mathrm{int}}\left(t\right)-\frac{i
\hbar}{2}\left(\kappa~\hat{a}^{\dagger}\hat{a}+\gamma~\hat{\sigma}_{+}\hat{\sigma}_{-}+\gamma_{\mathrm{deph}}\right)$,
where $\hat{a}$ is the cavity photon annihilation operator and
$\hat{\sigma}_{+},\hat{\sigma}_{-}$ are the exciton creation and
annihilation operators.
$H_{\mathrm{JC}}$ is the Jaynes-Cummings Hamiltonian and
$H_{\mathrm{int}}=\frac{\hbar\Omega\left(t\right)}{2}\left(\hat{a}+\hat{a}^{\dagger}\right)$
denotes the interaction with the Gaussian laser pulses,
$\Omega\left(t\right)=\Omega_{\mathrm{control}}~\mathrm{exp}\left(-i\omega_{\mathrm{control}}t-2~\mathrm{ln}\left(2\right)t^{2}/~T_{\mathrm{pulse}}^{2}\right)+\Omega_{\mathrm{signal}}~\mathrm{exp}\left(-i\omega_{\mathrm{signal}}t-2~\mathrm{ln}\left(2\right)\left(t+\tau\right)^{2}/~T_{\mathrm{pulse}}^{2}\right)$.
$\kappa$~denotes the cavity dissipation rate, $\gamma$ the exciton
spontaneous recombination rate and $\gamma_{\mathrm{deph}}$ the
exciton pure dephasing rate. $\Omega_{\mathrm{control}}$ and
$\Omega_{\mathrm{signal}}$ correspond to the maximal laser-cavity
coupling rates at the peak power of control and signal pulses, with
an overall $2~\%$ system coupling efficiency included.
$N_{\mathrm{control}}$ and $N_{\mathrm{signal}}$ are calculated in
the same way, with $\Omega_{\mathrm{signal}} = 0$ or
$\Omega_{\mathrm{control}} = 0$ respectively. In order to determine
$I\left(t\right)$, we use a Monte Carlo wavefunction (MCWF)
approach, as described in \cite{Molmer:JOSA93}. The calculations are
performed with the experimentally determined values,
$\lambda_{\mathrm{exciton}}=937.25~\mathrm{nm}$, $\hbar
g=141~\mathrm{\mu eV}$, $\hbar\kappa=53~\mathrm{\mu eV}$,
$\hbar\gamma=0.66~\mathrm{\mu eV}$, and
$\hbar\gamma_{\mathrm{deph}}=13~\mathrm{\mu eV}$
\cite{Reinhard:arXiv11}.

\subsection*{Cross-correlation of upper and lower polaritons}

In the main text, we demonstrate a positive non-linear response of
the system when driving the fundamental upper polariton transition
with a control pulse and the transition from the first to the second
Jaynes-Cummings manifold with a signal pulse. For pulse durations of
$T_{\mathrm{pulse}} > 50~\mathrm{ps}$, the width of the
delay-dependent non-linearity corresponds to a very good
approximation to the convolution of the original control and signal
pulses. Thus, the system can serve as a single-photon pulse
correlator. Similarly, one can use the two polariton transitions for
demonstration of pulse-correlator operation. We carried out the
corresponding experiment by exciting the upper polariton transition
with the control laser and the lower polariton with the signal
laser. In this case, the optical non-linearity is negative (compare
Figure 2c of the main text), since the ground state population is
decreased with the control/signal laser present, preventing the
other laser to scatter photons on the lower/upper polariton
transition, respectively. Without post-filtering the scattered
photons (corresponding to the situation in our experiment), the
control and signal pulses are inter-changeable without having any
effect on the non-linear system response. The negative non-linear
response on the polaritons is about a factor of two larger than the
positive non-linearity on the transition into the second
Jaynes-Cummings manifold, as can be seen in Fig.~2c of the main
text. Figure~S2 displays corresponding measured negative non-linear
system responses as a function of pulse delay for pulse durations of
(a) $33~\mathrm{ps}$ and (b) $51~\mathrm{ps}$.

\begin{figure}
\includegraphics[width= 84 mm]{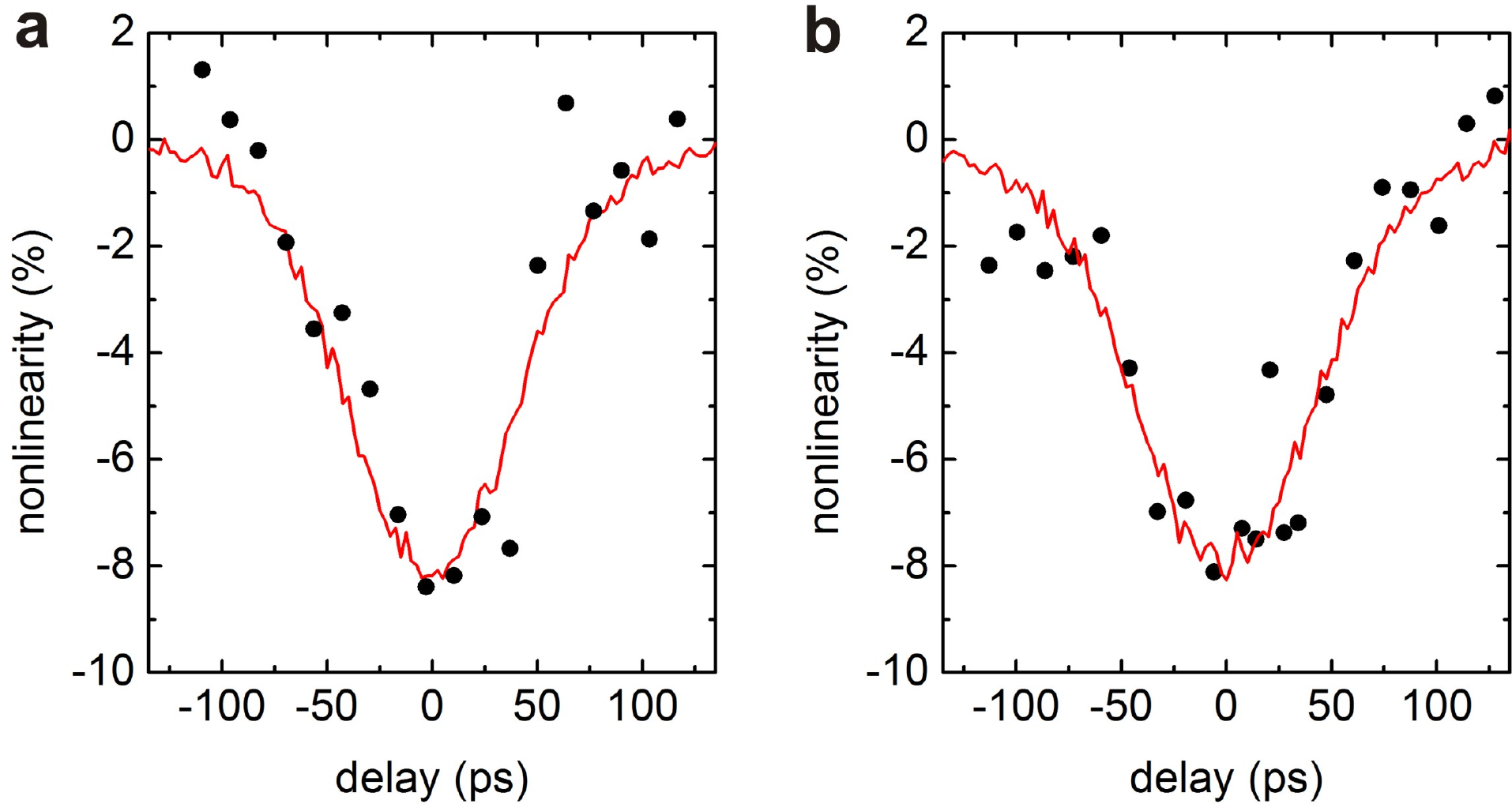}
\caption{{\bf Ultrafast negative non-linear system response.}
Non-linear system response of the QD-cavity device as a function of
pulse delay with the control laser set to the upper polariton and
the signal laser on the lower polariton transition for pulse
durations of {\bf a,} 33~ps and {\bf b,} 51~ps. Both lasers had an
average power of 1~nW. Since here the role of control and signal
lasers are interchangeable, the system response is symmetric. The
red curves are results from MCWF simulations with the absolute
amplitude determined from a least-square fit to the data.
\label{FigS2}}
\end{figure}

\subsection*{Transistor operation}

As suggested in Ref.~\cite{Faraon:NatPhys08}, the present device
could in principle be operated as a single-photon transistor. In
analogy to a conventional electronic transistor, a single-photon
transistor should exhibit gain. Following
Ref.~\cite{Chang:NatPhys07}, at the level of single light quanta one
can define the gain $G$ as the number of scattered signal photons
$N_s$, per incident control photon before the spontaneous relaxation
of the device back into its ground (or off) state. In a three-level
lambda system as considered in that article, $G$ is given by the
ratio of relaxation rates (or branching ratio) of the signal and the
control transition. Following this argument, one might naively
expect that in a Jaynes-Cummings system the gain should be
determined by the ratio of the excited state lifetimes of the first
compared to the second manifold. However, due to the finite overlap
of transitions between higher-lying manifolds, multiple transitions
are involved in the system dynamics. If $g/\kappa>2$, the
fundamental polariton transitions are to a certain extent spectrally
separated from the rest, i.e. from the higher-lying transitions. If
a weak control laser pulse is resonant with a transition from the
ground state to a polariton state while a stronger signal pulse
addresses the higher-lying transitions jointly, a gain larger than
one is possible due to the decreasing anharmonicity for higher
photon numbers. In a simple simulation using the MCWF method, we
find that a Jaynes-Cummings system with $g/\kappa = 2.7$ (our
device) can have a gain of at least $G\approx 2.2$ and an on/off
ratio of $\approx 5$, provided pure dephasing is neglected. If
$g/\kappa$ is increased by a factor of 4 (as could be achieved with
larger Q in state-of-the-art nano-cavities), a gain of $G\approx
6.5$ and an on/off ratio of $\approx 13$ could be obtained. The
calculated gain rapidly decreases for the experimentally observed
dephasing rates. Hence, a clear experimental demonstration of
transistor gain appears to be very demanding with the current system
parameters.

\end{document}